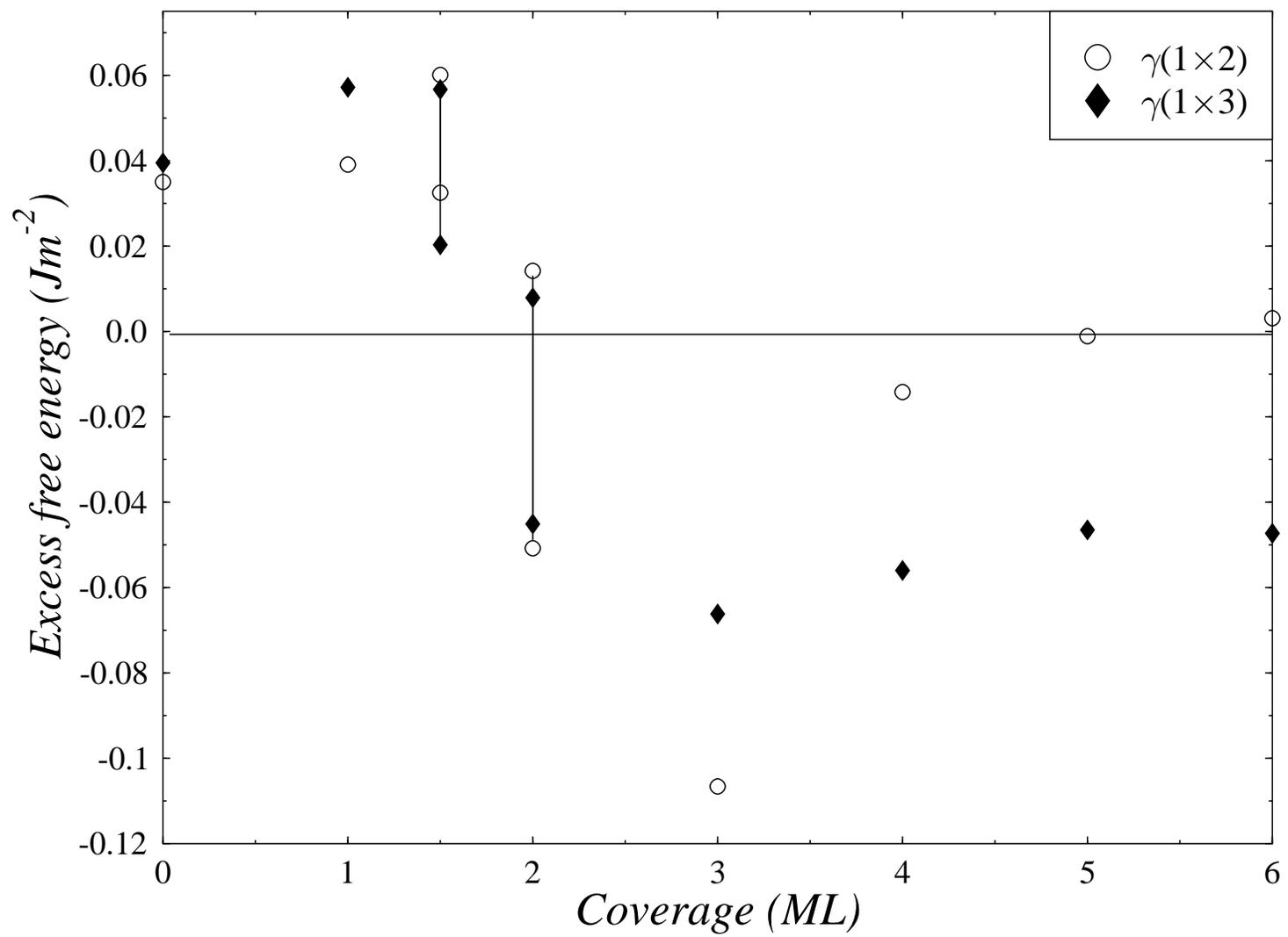

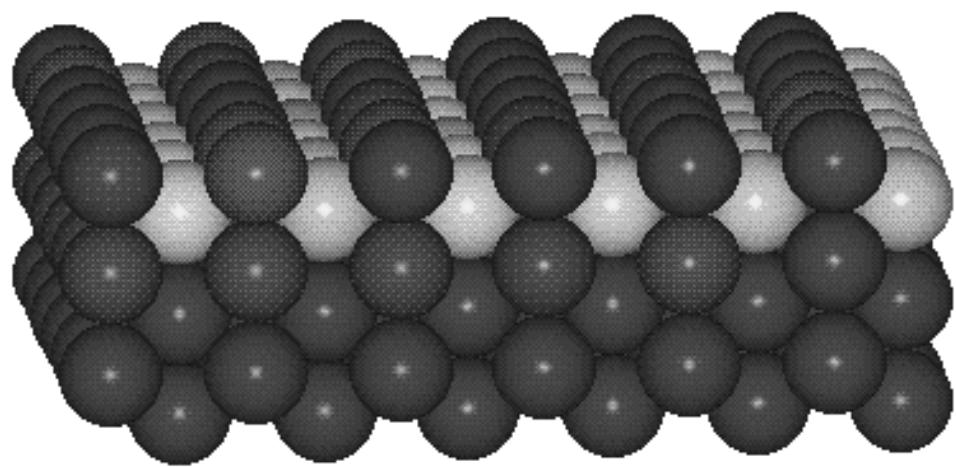

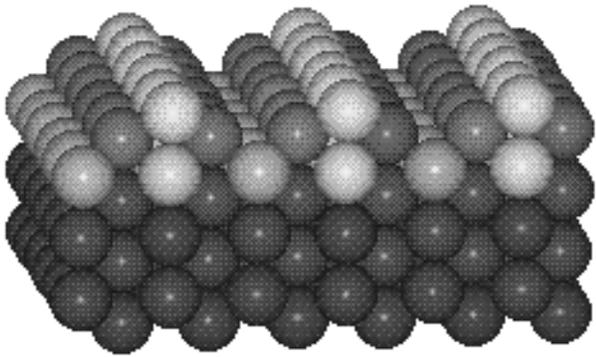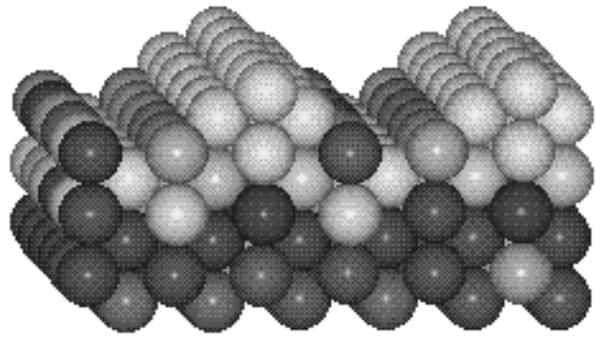

# Atomic exchange mean field study of intermixing of Au on Ag(110)

J. A. Nieminen

*Department of Physics, Tampere University of Technology, P.O.Box 692,*

*FIN - 33101 Tampere, Finland*

*E-mail: Jouko.Nieminen@cc.tut.fi*

## Abstract

A mean field method to study heteroepitaxial thin film growth is applied to growth, intermixing and surface reconstructions of $Au$ on $Ag(110)$. The results are in accordance with experimentally observed "burrowing" at submonolayer coverages and growth of elongated, $(1 \times 3)$ reconstructed, $Au$ clusters at higher coverages. At coverages of few monolayers the surface between the clusters has a high concentration of $Ag$, and ordered rows of $Au$ are formed just beneath the surface.

Studying heteroepitaxial growth on (110) surface of noble and transition metals is a challenge not only to experimental surface scientists, but the related problems of structural phase transitions and interdiffusion are also of a more general interest in statistical and computational physics of condensed matter. Regarding pure fcc $d$-metals, the $4d$-metals, $Rh$, $Pd$ and $Ag$, have a stable unreconstructed (110) surface, whereas $5d$-metals, $Ir$, $Pt$ and $Au$, favour spontaneous $(1 \times 2)$ missing row reconstruction [1]. This behaviour can be understood either in terms of the relative contribution of $d$-electrons to binding [2] or in terms of the range of interatomic potentials [3,4]. $Au$ on $Ag(110)$ [5–10] and $Au$ on $Pd(110)$ [11–13] are examples of heteroepitaxial $d$-metal systems which have inspired many interesting experimental and theoretical studies within the last five years. Predicting the behaviour of such systems is nontrivial since their compounds have different reconstruction behaviours, are soluble to each other, and may have a considerable lattice mismatch.

The growth behaviour of $Au$ on $Ag(110)$ at room temperature is rather unexpected because of a coverage dependent intermixing of the compounds. The Scanning tunneling microscopy (STM) studies by Rousset et al. [7] combined with a reinteptretation of Medium-energy ion scattering (MEIS) studies by Fenter and Gustafsson [5,6] show that at submonolayer coverages $Au$ interdiffuses or "burrows" into the substrate and a monolayer of $Ag$ is formed on the top of the $Au$ film. At coverages above $1ML$ $Au$ does not prefer interdiffusion but $(1 \times 3)$ reconstructed $Au$ "fingers", or clusters elongated in $[1\bar{1}0]$ direction, are formed on the top of $Ag$. The first theoretical confirmation of the interdiffusion picture was given by the total energy calculations by Chan et al [8]. Using Molecular dynamics (MD) simulations Haftel et. al. [9] were able to describe the dynamics of the atomic replacement mechanism of the interdiffusion at submonolayer coverages and the formation of clusters elongated in $[1\bar{1}0]$ direction at higher coverages, thus supporting the description of the growth mode given in Ref. [7].

MD simulations model the real dynamics of interdiffusion at finite temperatures, and are not restricted to any predefined structure, which are great advantages of the method. However, MD simulations suffer from the existence of high activation barriers between different configurations which slow down diffusion. Haftel et. al. [9] have succesfully utilized an MD simulation analogous to simulated annealing (compare to Ref. [14]) to avoid this problem: they start from a high temperature to make the atoms sufficiently mobile and slowly cool the system down to a lower temperature in order to find the equilibrium structure. This approach gives the overall picture of the growth mode right, but the high starting temperature possibly creates disorder that gets "frozen" as the system is cooled down. On the other hand, the surface structure may depend on temperature in a complicated way, as is seen in Refs. [11–13] for $Au$ on $Pd(110)$.

In this Letter, the growth of $Au$ on $Ag(110)$ is studied using a *mean field* theory for minimizing the mixing *free energy* with respect to the atomic configuration and local spatial relaxations. The present approach searches for energetically the most favourable *equilibrium* structure of the surface for different coverages, whereas it does not attempt to simulate *kinetics* of the structural ordering. The free energy formulas by Najafabadi et. al. for segregation of compounds $A$ and $B$ of binary alloys at grain boundaries [15] are used as a starting point. In that model the *Grand potential*, $\Omega$, has contributions from internal energy and atomic vibrations, $F_v$, accompanied by mixing entropy, $S$:

$$F_v = E_P + 3kT \sum_i \ln[\frac{\hbar |D_i|^{1/6}}{\sqrt{M_i kT}}] \tag{1a}$$

$$\Omega = F_v + k_B T \sum_i [c_i ln(c_i) + (1-c_i) ln(1-c_i) + (\mu_A - \mu_B) c_i], \tag{1b}$$

where $E_P$ is the potential energy and $|D_i|$ is the determinant of the $3 \times 3$ matrix $D_{i\alpha\beta} = \frac{\partial^2 E_P}{\partial u^i_\alpha \partial u^i_\beta}$ from the Einstein approximation to local quasiharmonic approximation of the atomic vibrations [16]. In the entropy term the parameter $c_i$ gives the probability of the atom $i$ to be of the species $A$, and if no vacancies are assumed, the probability of the species $B$ is $1 - c_i$.

The form of potential energy can be specified by using, e.g., Sutton-Chen (SC) potential in a mean field form [17]:

$$E_P(\{c_i\}, \{\bar{r}_i\}) = \frac{1}{2} \sum_{i,j \neq i} [c_i c_j V^{AA}_{ij} + (1-c_i) c_j V^{AB}_{ij} + (1-c_i)(1-c_j) V^{BB}_{ij}]$$

$$-\sum_i [c_i \epsilon^{AA} d^A + (1-c_i)\epsilon^{BB} d^B]\sqrt{\rho_i}, \qquad (2)$$

where

$$\rho_i = \sum_{j \neq i}[c_i c_j \phi_{ij}^{AA} + (1-c_i)c_j \phi_{ij}^{AB} + (1-c_i)(1-c_j)\phi_{ij}^{BB}],$$

$V_{ij}^{\alpha\beta} = \epsilon^{\alpha\beta}(\frac{a_{\alpha\beta}}{r_{ij}})^{n_{\alpha\beta}}$ and $\phi_{ij}^{\alpha\beta} = \epsilon^{\alpha\beta}(\frac{a_{\alpha\beta}}{r_{ij}})^{m_{\alpha\beta}}$. The potential parameters $\epsilon, a, m, n$ and $d$ are as described in Refs. [18,19]. It is easy to apply SC potential to Eq. (1), in practice, and there are simple but satisfactory rules to model bonding between different species [19]. The potential gives the correct reconstruction behaviour for the (110) surfaces of, e.g., $Pd$, $Ag$, $Pt$ and $Au$ [3,13]. As combined to a local Einstein approximation to the quasiharmonic theory SC has also given a qualitatively correct temperature and coverage dependent structural phase diagram for $Au$ on $Pd(110)$ [13].

The mean field model above has been applied to interfaces [15,17] assuming infinite reservoirs of both species (*Grand canonical ensemble*) and allowing the difference between chemical potentials of the species govern the local concentration of the species at the interface. This approach is not useful in the case of a fixed number of adsorbate atoms where the ensemble is *canonical*. The optimization in the canonical case must take place as exchange of concentration between neighbouring atoms to model interdiffusion by atomic exchange mechanism. This approach is analogous to spin conserving *Canonical Monte Carlo* methods, in contrast to the ordinary *Grand Canonical Monte Carlo* [20], which corresponds to the mean field method of Refs. [15,17]. From here on the present method is called atomic exchange mean field (AEMF) method.

In AEMF the Grand potential, $\Omega$, is optimized with respect to $c_i$ constrained to conserve the total number of both $Au$ and $Ag$ atoms. The exchange of concentration is allowed by choosing, in random order, an atom $p$ and one of its neighbours, $q$. In equilibrium, the balance between the two neighbours is expressed as $\frac{\partial \Omega}{\partial c_p} - \frac{\partial \Omega}{\partial c_q} = 0$, or explicitely from Eq. (1):

$$c_p = \phi(c_p, c_q) \equiv \frac{c_q}{c_q + (1-c_q)\exp[-F_{p,q}/kT]}$$

where
$$F_{p,q} = \frac{\partial F_v}{\partial c_p} - \frac{\partial F_v}{\partial c_q}.$$

It should be noticed that the balance condition cancels the difference in chemical potentials, $\mu_A - \mu_B$. The constrained optimum could be searched by applying the equilibrium condition iteratively for randomly chosen pairs of atoms, but a better convergence is obtained when the rate of the concentration exchange is moderated according to

$$c_p^{new} = \beta(\alpha c_p^{old} + \phi(c_p^{old}, c_q^{old}))$$
$$c_q^{new} = \beta(\alpha c_q^{old} + \phi(c_q^{old}, c_p^{old})) \quad (3)$$

where $\beta$ is a normalization constant to conserve $c_p + c_q$. In most of the cases $\alpha = 4$ was found to give a good convergence in about 100 iterations per a pair of atoms, but also higher values of $\alpha$ were used [21]. In intervals of a few iterations of Eq. (3), also *local* spatial relaxations were allowed using a variant of Polak-Ribiere optimization method [22] to seek the minimum of $\Omega$ with respect to $\{\bar{r}_i\}$.

The calculations were carried out at room temperature using a slab of 22 layers in [110]-direction, each layer consisting of 6 rows (in [1$\bar{1}$0]-direction) and 6 columns (in [100] direction) with periodic boundary conditions in these directions. The minimum size of the calculational cell is limited by the cut-off radius of the SC potential, $r_c = 2a$, and the number of rows must be divisible by 2 and 3 for $(1 \times n)$ reconstructions with $n$ either 2 or 3 and the slab must be thick enough to model substrate bulk properties deeper within the slab. In the vertical [110]-direction eight layers at the very bottom of the slab have fixed positions and concentrations, $c_i = 1.0$, in order to maintain $Ag$ bulk structure. The species $A$ of Eq. (2) is now $Ag$ and $B$ is $Au$. The *initial* concentrations $c_i$ of $Ag$ are given as follows: a predefined number of surface layers is set to be of $Au$ and the rest of the atoms in the slab have $c_i = 0.9999$ to avoid overflow in calculating the entropy. The concentration is optimized according to Eq. (3) and local relaxations are done to all the atoms except the eight layers at the very bottom of the slab. The surface free energy is calculated by subtracting the free energy per atom of $Ag$ bulk from the free energies of the atoms of the surface layers (in this case 14 layers) and by dividing the total sum by the surface area similarly to, e.g., Refs. [3,4,13].

I have calculated the free energies for optimized concentration profiles of the $(1 \times 1)$, $(1 \times 2)$ and $(1 \times 3)$ structures for various coverages at $\Theta = 0 - 6ML$. As seen in Fig. 1, there is a strong coverage dependence of the relative free energies of different surface structures for optimal concentration profiles. A good convergence and clear results are obtained at submonolayer coverages and coverages at and above $3ML$. At submonolayer coverages the surface remains unreconstructed (Fig. 1) and burrowing of $Au$ is seen very clearly (Fig. 2). In equilibrium at $\Theta = 1ML$ the top layer is virtually of pure $Ag$ and the $Au$ concentration of the second layer is 85%. When increasing the coverage for the unreconstructed structure, further interdiffusion is not favourable, but $Au$ atoms tend to stay in the top layers. The increasing concentration of $Au$ in the top layer makes the $(1 \times 1)$ structure less favourable. Fig. 1 also shows that the $(1 \times 2)$ structure would be the most favourable for $\Theta = 2ML$ and $3ML$, since that structure has the best, although a remote, resemblance to the experimentally observed small clusters. However, the poor convergence around $\Theta = 2ML$ means that the predefined structure is not very satisfactory at that coverage.

Above $\Theta = 3ML$, the $(1 \times 2)$ structure grows increasingly unfavourable, and the $(1 \times 3)$ structure is about $0.05 Jm^{-2}$ below the free energy of the $(1 \times 1)$ structure. This is consistent with the observed $(1 \times 3)$ reconstruction and elongated islanding in $[1\bar{1}0]$ direction. Furthermore, this resembles the observation by Fenter and Gustafsson [6], who saw a broad $(1 \times 1) \rightarrow (1 \times 3)$ transition above $\Theta \approx 3.6ML$, which suggests a gradual formation of elongated clusters. If the $(1 \times 2)$ structure is assumed at $\Theta \approx 2ML$, a certain ordering of $Au$ atoms at the surface occurs. $Au$ tends to form clusters virtually free from $Ag$, such that the top layer and one row immediately below the top row is of $Au$. The two rows in the second layer between the clusters are of almost pure $Ag$, while there is a single $Au$ row below the two $Ag$ rows (see Fig. 3). This is probably the mechanism of growing clusters on the surface, since the same pattern is seen also for the $(1 \times 3)$ structure at $\Theta = 2ML$ and at higher coverages, as well. Fig. 3 shows the configuration at $\Theta = 4ML$, where the $Ag$ concentration of the two rows between the clusters is lower than at lower coverages but does not vanish. A further increase of the coverage makes $Ag$ completely to vanish from the surface. It also increases interdiffusion which makes the interface between $Au$ film and $Ag$ substrate very diffuse.

The ordered atomic configurations at coverages of few monolayers are not seen in the MD simulations by Haftel. et. al [9] although they otherwise give overall results similar to the present calculations. The difference arises obviously from high mobility of atoms at high temperatures in MD simulations and a possible frozen disorder at lower temperatures. It should be emphasized that MD gives unique momentary configurations of atoms comparable to STM observations, whereas the ordered configurations given by AEMF are statistical averages accessible to scattering experiments. Thus, the ordered configurations shown in Fig. 3 could possibly be seen experimentally, e.g., by analyzing LEED intensities.

A qualitative explanation of the present results is found from the properties of the semiempirical SC-potential. At *room temperature* SC gives surface and interface free energies $\gamma_{Au,(1\times2)} = 0.55 Jm^{-2}$, $\gamma_{Ag} = 0.98 Jm^{-2}$ (consistent with values in Refs. [3,4,23]) and $\gamma_{Au-Ag} = -0.07 Jm^{-2}$, the latter for a sharp interface. The cohesive energies are $E_{Au-Au} = -3.78 eV$, $E_{Ag-Ag} = -2.96 eV$ and $E_{Au-Au} < E_{Au-Ag} < E_{Ag-Ag}$ the value of $E_{Au-Ag}$ depending on the environment. Thus, $\gamma_{Au} - \gamma_{Ag} - \gamma_{Au-Ag} = -0.36 Jm^{-2} < 0$, which would strongly favour forming of ($(1\times2)$ reconstructed) $Au$ film on the top of the $Ag$ surface. However, due to long range of interatomic potentials this argument is valid only for thick $Au$ films. A word of caution should also be said due to Ref. [4], which shows that very different surface energies, $\gamma$, for $Ag$ and $Au$ are obtained by different methods (EMT, EAM, TB or Experimental). At low coverages, interlayer interactions $V_{Au-Au} < V_{Au-Ag} < V_{Ag-Ag}$ can be assumed, reflecting the corresponding cohesive energies. The energy of burrowing one $Au$ layer below an $Ag$ layer is $\Delta V = V_{Au-Ag} - V_{Ag-Ag} < 0$ indicating that a *very thin* $Au$ film on the top has a *high* surface energy. For coverages at and below one monolayer the surface energy is decreased by burrowing, but at higher coverages it is possible to cover open $Ag$ surface simultaneously with increasing coordination of $Au$ atoms by forming clusters of $Au$. Thus it is the competition of these two mechanisms to decrease the free energy of the film that lead to a cross-over in the growth mode.

To conclude, the AEMF gives results consistent with recent experiments and theoretical calculations for $Au$ on $Ag(110)$. In addition, it produces predictions about detailed atomic configurations, which may be experimentally confirmed. Although AEMF has been applied to a specific system here, it is also available to other heteroepitaxial systems with different structures and compounds, taken that suitable interatomic potentials exist. A major improvement to the method would be self-optimization of the surface structure, which would allow formation of clusters with an arbitrary size and shape.

**Acknowledgements:** Dr. A.P. Sutton is very gratefully acknowledged for giving useful hints concerning free energy calculations and the mean field method.

FIGURES

FIG. 1. The surface free energies for optimal configurations of the $(1\times 2)$ and $(1\times 3)$ structures relative to the surface free energy of the $(1 \times 1)$ structure at different coverages. The markers connected by lines show the error limits at coverages where the convergence was not satisfactory.

FIG. 2. Optimal atomic configuration for $\Theta = 1ML$ assuming the $(1 \times 1)$ structure. The shade of the balls show the concentrations, $c_i$, so that light shades denote low values of $c_i$, i.e., a high probability of $Au$.

FIG. 3. Optimal atomic configurations for $\Theta = 2ML$ with the $(1 \times 2)$ structure (left) and $\Theta = 4ML$ with the $(1 \times 3)$ structure (right). The shades of the balls have the same meaning as in the previous figure.